\documentclass[12pt,preprint]{aastex}
\def\FIR{\ifmmode {\,$\tau_{\rm FIR}$} \else $\,\tau_{\rm FIR}$\fi}
\def\tk{\ifmmode {\,$T_{\rm k}$} \else $\,T_{\rm k}$\fi}
\def\mic{\ifmmode {\,\mu{\rm m}} \else $\,\mu{\rm m}$\fi}       % micron
\def\kms{\ifmmode {\,{\rm km\,s^{-1}}}                          % km s-1
        \else {\hbox{$\,$ {\rm km$\,$s$^{\rm -1}$}}}\fi}
\def\mo {\ifmmode {\,{\it M}\odot} \else $\,M$\odot\fi} % M solar
\def\lo {\ifmmode {\,{\it L}\odot} \else $\,L$\odot\fi} % L solar
\def\my {\ifmmode {\,{\it M}\solar\,{\rm yr^{-1}}}              % Msol/year
        \else {$\,M$\solar$\,$yr$^{\rm -1}$}\fi}
\def\cmm#1{\ifmmode {\,{\rm cm^{-#1}}}          % cm-1, cm-2, cm-3,....
        \else \hbox{$\,${\rm cm$^{\rm -#1}$}}\fi}
\chardef\isp="10\def\i{\'\isp}          % dotless i withaccent
\def\as {\ifmmode {^{\scriptscriptstyle\prime\prime}}           % arcsec
        \else $^{\scriptscriptstyle\prime\prime}$\fi}
\def\am {\ifmmode {^{\scriptscriptstyle\prime}}                 % arcmin
        \else $^{\scriptscriptstyle\prime}$\fi}
\def\deg {\ifmmode^\circ\else$^\circ$\fi}                       % degree
\def\raw {\ifmmode\rightarrow\else$\rightarrow$\fi}             % rightarrow
\def\x {\ifmmode\times\else$\times$\fi}                         % times (x)
\def\gsim {\ifmmode {\buildrel>\over\sim}               % greater or similar
        \else {\lower.6ex\hbox{$\buildrel>\over\sim$}}\fi}
\def\lsim {\ifmmode {\buildrel<\over\sim}               % less or similar
        \else {\lower.6ex\hbox{$\buildrel<\over\sim$}}\fi}
\def\ra[#1 #2 #3.#4]{ #1$^{\rm h}$#2$^{\rm m}$#3$^{\rm s}$.#4}  % RA
\def\dec[#1 #2 #3.#4]{ #1\deg#2\am#3{\as}.#4}             % declination
\def\rax[#1 #2 #3]{RA: #1$^{\rm h}$#2$^{\rm m}$#3$^{\rm s}$}%RA with
\def\decx[#1 #2 #3]{Dec:#1\deg#2\am#3\as}          % Dec with no fraction
\def\HII{{\sc Hii}}                   % HII
\def\uch{UC\,{\sc Hii} region}           % UC HII region
\def\h2{\rm H$_2$}                      % molecular hydrogen
\newcommand{\cop}{CO$^+$}               % CO(+)
\newcommand{\hocp}{HOC$^+$}             % HOC(+)
\newcommand{\g}{G29.96-0.02}
\def\apj{{\rm ApJ}}                     % Astrophys. J.
\def\apjs{{\rm ApJS}}                   % Astrophys. J. Supl.
\def\apjl{{\rm ApJ Let}}                % Astrophys. J. (Letters)
\slugcomment{Not to appear in Nonlearned J., 45.}

\shorttitle{Reactive ions in UC\, {\sc Hii} regions}
\shortauthors{Rizzo et al.}

\begin{document}


\title{Detection of reactive ions in the ultracompact
{\sc Hii} regions Mon R2 and G29.96-0.02}


\author{J. R. Rizzo\altaffilmark{1,2}, A. Fuente\altaffilmark{2}, 
A. Rodr{\i}guez-Franco\altaffilmark{3} and 
S. Garc{\i}a-Burillo\altaffilmark{2}}


\affil{$^1$ Departamento de F{\i}sica, Universidad Europea de Madrid, 
Urb.\ El Bosque, E-28670 Villaviciosa de Od\'on, Spain\\
       $^2$ Observatorio Astron\'omico Nacional, Aptdo. Correos 1143,
 E-28800 Alcal\'a de Henares, Spain\\
       $^3$ Dpto. de Matem\'atica Aplicada, Universidad Complutense de Madrid, 
Av. Arcos de Jal\'on s/n, E-28037 Madrid, Spain}

\email{jricardo.rizzo@fis.cie.uem.es, a.fuente@oan.es, 
arturo@damir.iem.csic.es, s.burillo@oan.es}


\begin{abstract}
We report the first detection of the reactive ions \cop\ and \hocp\ 
towards ultracompact (UC) \HII\ regions, particularly in Mon R2 and 
G29.96-0.02. We have observed two positions in Mon R2, namely the peak 
of the \uch\ and the surrounding high density molecular cloud. \cop\ 
and \hocp\ were detected at the \uch\ but not at the 
molecular cloud, as expected if both ions arise 
in the PDR surrounding the \uch. The measured column 
densities are of the order of 10$^{11}$ cm$^{-2}$ in both sources, which
yields a strikingly low [HCO$^+$]/[HOC$^+$] abundance ratio of 460 in 
Mon R2. These values are similar to those found in some other well-known 
PDRs, like NGC 7023 and the Orion Bar. We briefly discuss the chemical 
implications of these results.
\end{abstract}




\section{Introduction}

The chemistry of PDRs is nowadays the subject of an important
amount of observational and theoretical works.
The reactive ions are among the best tracers of PDRs, because their abundances
are expected to be negligible in the shielded parts of molecular clouds. 
Actually, these compounds react on virtually every collision with H$_2$ and
only in the hot layers of PDRs, where a significant
fraction of hydrogen is still in atomic form, does the abundance of these
ions become significant (Sternberg \& Dalgarno 1995). Recently, our group
has carried out a survey of reactive ions (CO$^+$, HOC$^+$, SO$^+$)
towards the prototypical PDR regions NGC 7023, the Orion Bar and the
planetary nebula NGC 7027 (Fuente et al.~2003).
Our results confirm that the abundance of reactive ions is strongly
enhanced in PDRs, although strong differences may exist among PDRs depending
on their physical and chemical conditions.

So far, all the observational knowledge about reactive ions in PDRs are based
on a few sources. Very little is known about the  chemistry of the
PDRs associated to ultracompact (UC) \HII\ regions.
The high densities ($>$ 10$^6$ cm$^{-3}$) and kinetic temperatures
($\sim$ 100 K) of these regions, together with the
ionizing radiation (G$_0 > 10^5$ in units of Habing field), could drive a
chemistry presumably different from that found in other PDRs. 
In this paper we present the detection of CO$^+$ and HOC$^+$
towards the \uch s Mon R2 and G29.96-0.02. These are the first detections 
of CO$^+$ and HOC$^+$ in this kind of regions and prove the feasibility of 
chemical studies of the PDRs associated with \uch s, even at large
distances, as in the case of \g\ (6 or 9 kpc, Pratap et al.\ 1999).

\section{Observations}

We have carried out a survey of the reactive ions \cop\ and \hocp\ towards 
the PDRs associated to the \uch s Mon R2 and G29.96-0.02. Besides \cop\ 
and \hocp, we have also observed some rotational transitions of SiO, 
H$^{13}$CO$^+$ and HC$^{18}$O$^+$. The observations were carried out using 
the IRAM 30m radio telescope in Pico de Veleta (Spain) during July 2002. 
The observed transitions, their frequencies and the telescope parameters are 
shown in Table 1.
%The HOC$^+$ and HC$^{18}$O$^+$ 1$\rightarrow$0 lines have always been 
%observed simultaneously, in order to obtain a more accurate 
%[HC$^{18}$O$^+$]/[HOC$^+$], and avoiding the uncertainties due to 
%calibration and pointing. We also observed simultaneously the SiO 
%2$\rightarrow$1 and 3$\rightarrow$2 lines, which allowed us to
%estimate the hydrogen densities in the studied regions.

\section{Mon R2}
Mon R2, located at a distance of 950 pc, is the closest \uch\
and has an angular diameter of $\sim$ 22$''$ (0.1 pc).
It is comparable to the beam of the 30m radio telescope
at 3mm, which allows us to spatially
separate the \uch\ and the surrounding molecular cloud.
The situation is clearly illustrated in Fig.~1. In the left panel, the CS
7$\rightarrow$6 emission map from Choi et al.~(2000) is shown. The central 
square roughly indicates the angular extension of the \uch\ at 
6cm-continuum (Wood \& Churchwell 1989). The beamsize both
at 3mm and 1mm are also sketched. The CS emission
shows an arc-like structure surrounding the \HII\ region, with the
maximum of emission to the southeast. The \uch\ is highly asymmetric and has
a cometary shape, reaching its maximum toward its exciting star Mon R2 IRS1,
which is also the origin of angular offsets. We have performed
the observations towards this maximum of the ionized gas, at (0\arcsec,
0\arcsec), and toward the peak of CS emission, at (10\arcsec, -10\arcsec).
Some of the observed spectra at both positions are shown in the right panel
of Fig.~1. While the HC$^{18}$O$^+$ emission is more intense in the molecular
cloud than in the \HII\ region position, the reactive ions
CO$^+$ and HOC$^+$ have only been detected towards the \uch.
Both CO$^+$ lines have been detected towards the \uch, and fulfill the 
line ratio of 0.55 expected for optically thin emission.

The parameters of the Gaussian fits to all the observed lines at the two 
positions in Mon R2 (and also the single position in \g) are shown
in Table 2. The H$^{13}$CO$^+$ and HC$^{18}$O$^+$ lines seem to mimic the CS
J = 7$\rightarrow$6 line emission --a well known tracer of high density gas--,
because both lines are enhanced in the molecular cloud
with respect to the \HII\ region. This enhancement is indeed a lower limit,
because these lines were observed with a 29\arcsec\ beam.
However, the behaviour of CO$^+$ and HOC$^+$ is the opposite.
Since the dipole moments of both ions are similar to that of HC$^{18}$O$^+$ and
H$^{13}$CO$^+$, this effect may not be due to an excitation effect.
Actually, the detection of these ions towards the ionized region and the
lack of detection towards the molecular envelope can only be understood
if the abundances of CO$^+$ and HOC$^+$ are strongly enhanced toward
the position of the \uch\ relative to those in the molecular cloud.
Very likely, the emitting region of these ions would be the PDR associated
to Mon R2.

\section{\g}

The situation in \g\ is more complex  than in Mon R2.
Besides a surrounding massive molecular envelope, there is also a hot core 
at 2\arcsec\ from the center of the \uch. So far, our single-dish 
observations were done toward a single position which includes, even at 1mm, 
contributions from the \uch, the molecular envelope and the hot core. 
The positions of 
\g\ and its neighbour hot core, as well as the relevant spectra observed 
in this paper, are shown in Fig.~2.

Both the hot core and the dense envelope associated with this source
have been extensively observed in molecular lines (see e.g.~Cesaroni et 
al.~1998; Maxia et al.~2001).
Fortunately, we can identify and separate the emission of the hot core
by kinematics considerations. While the lines arising in the hot core should
have linewidths $\sim$ 15--20 \kms, the lines arising in the molecular cloud
and/or PDR should have linewidhts $\sim$ 3--5 \kms. In Fig.~2 (lower left
panel), the SiO 3$\rightarrow$2
and  SiO 2$\rightarrow$1 spectra are shown. We see that these spectra have
a narrow and a wide components. While the narrow component
arises from the \uch\ and the molecular environment, the wide component
arises from the hot core.

Figure 2 (right panel) shows the HC$^{18}$O$^+$, CO$^+$ and HOC$^+$ 
spectra observed in \g. The \cop\ lines does not fulfill the standard 
ratio of 0.55, because of 
%since the CO$^+$ 2,5/2$\rightarrow$1,3/2 line is wider 
%and several times more intense than the 2,3/2$\rightarrow$1,1/2 line. 
%This is due to 
the blending of the 2,5/2$\rightarrow$1,3/2 line by two hyperfine 
components of $^{13}$CH$_3$OH, which is expected to be abundant in the 
hot core and the molecular cloud. The contamination was confirmed by
the broad bandwidth (256 $\times$ 1 MHz-filterbank) spectrum, 
which shows the ``forest'' of 
$^{13}$CH$_3$OH, CH$_3$CHO and HCO$_2$CH$_3$ lines (Blake et al.~1984).
In order to avoid contamination with the $^{13}$CH$_3$OH lines,
we have used the CO$^+$ 2,3/2 $\rightarrow$ 1,1/2 line for our
column density estimates.

Both the CO$^+$ line and the HOC$^+$ line are narrow (up to a few \kms) and
hence they are not originated in the hot core. Specially striking is the
double-peaked structure of the 2,3/2$\rightarrow$1,1/2 line. The main peak, 
at $\sim$ 98.0 \kms,
has the velocity of the \uch, while the secondary peak, at $\sim$ 101.3 \kms,
is clearly redshifted. It is worth noting that there is HCO$^+$
emission also at this redshifted velocity (Maxia et al.~2001), as well as a
weak emission in our HC$^{18}$O$^+$ spectra. As discussed by Maxia et 
al.~(2001), the HCO$^+$ emission
is associated to the \uch. The simplest explanation for the redshifted
component is that we are seeing
a thin layer of dense gas in the receding face of the \uch. The fact
that both CO$^+$ peaks are similar in intensity, while the main part of the
HC$^{18}$O$^+$ spectrum is $\sim$ 10 times larger than the redshifted
part, implies that the CO$^+$ abundance would be $\sim$ 10 times larger in
the receding layer with respect to the 98 \kms\ gas. This shows that the
chemistry of this receding layer is being dominated by UV radiation from the
\uch. We are very likely seeing the innermost layer of molecular gas which 
is being accelerated by the ionized gas.

\section{CO$^+$ and HOC$^+$ as tracers of dense PDRs}

We have used the SiO J = $2\rightarrow 1$ and $3\rightarrow 2$ lines to
estimate the hydrogen densities in these regions. Using a LVG code and 
assuming for both sources the kinetic temperature (T$_{\rm k}$) of Mon 
R2 (50 K, Giannakopoulou et al.~1997), we have
derived hydrogen densities, $n$, of 1.5 10$^6$ cm$^{-3}$ for the
(0\arcsec,0\arcsec)  position of Mon R2, and 1.0 10$^6$ cm$^{-3}$ and
7.0 10$^5$ cm$^{-3}$ for the wide and narrow component of the emission
towards \g. These estimates are in reasonable agreement with other
species already observed, as referred above. 
The H$^{13}$CO$^+$ column density has been determined using 
the LVG code and
T$_{\rm k}$ = 50 K and $n$ = 10$^6$ cm$^{-3}$. With these assumptions, the 
H$^{13}$CO$^+$ rotation temperature is $\sim$ 20 K. Since the dipole 
moments of the observed reactive ions are similar to that of
H$^{13}$CO$^+$  we have derived the CO$^+$, HOC$^+$ and HC$^{18}$O$^+$
column densities assuming optically thin emission and using the LTE 
approximation with a rotation temperature of 20 K.
In Table 3, we show the column densities derived in this way for the
two positions observed in Mon R2 and for the narrow and wide component in \g.

Several arguments support the reactive ions CO$^+$ and HOC$^+$
arising in the PDRs linked to these \uch s. To start with,
a significant variation in the CO$^+$ abundance
can be found between
the molecular peak and the \uch\ in Mon R2 (see Table~3).
The [CO$^+$]/[HC$^{18}$O$^+$]
ratio is  at least a factor of 6 larger towards the \uch\  than towards
the molecular cloud.  Assuming a uniform HC$^{18}$O$^+$ abundance, this means
that the CO$^+$ abundance is at least a  factor of 6 larger in the \uch\
than in the  surrounding molecular cloud.
Like CO$^+$, the reactive ion HOC$^+$ has only been detected towards
the \uch\ in Mon R2. Assuming  $^{16}$O/$^{18}$O $\sim$ 650, we have
measured [HCO$^+$]/[HOC$^+$] $\sim$ 460 in Mon R2. This value is similar
to that found in the Orion Bar but
significantly lower than the values found by Apponi \& Ziurys (1997)
in a sample of star forming regions.

The CO$^+$ and HOC$^+$ column densities are
similar in both objects. This is quite surprising taking into
account that the column densities of the other molecules 
(H$^{13}$CO$^+$, HC$^{18}$O$^+$ and SiO) are at least one order of 
magnitude larger in \g\ than in Mon R2. It 
may be explained if CO$^+$ and HOC$^+$ arise in the PDR while the other 
species are well spread in the whole molecular cloud. Actually, this is an
indirect prove of the link of the reactive ions CO$^+$ and HOC$^+$ with 
the \uch.

\section{Discussion: \cop\ and \hocp\ chemistry}

The observation of reactive ions in regions with a wide range of physical
conditions is necessary to disclose between the different chemical
processes involved in the chemistry of reactive ions. Thus far, most of the
studied PDRs have low densities, $n$ $\leq$ 10$^5$ cm$^{-3}$, and incident
UV fields, G$_0$ $\leq$ 10$^5$. There is a lack of
data on regions of high densities and ionizing radiation. Mon R2 and \g\
are the first targets where the reactive ions CO$^+$ and HOC$^+$ can be
studied in such conditions.

We have estimated the incident UV field
by assuming that the observed FIR emission (L$_{\rm FIR}$) represents
all the stellar flux. Then a black body radiation law at the exciting 
star's effective temperature (T$_{\rm eff}$) is assumed to estimate 
the fraction of the luminosity radiated in the UV range. Adopting 
values for T$_{\rm eff}$, L$_{\rm FIR}$ and distance to the star of 
24000 K, 3000 L$_{\odot}$ and 0.03 pc for Mon R2 (Henning et al.~1992), 
and 35000 K, 45000 L$_{\odot}$ and 0.2 pc for \g\ (Morisset et 
al.~2002; Mart{\i}n-Hern\'andez et al.~2003), we have obtained 
G$_0 = 4.9\ 10^5$ for Mon R2 and G$_0 = 1.5\ 10^5$ for \g. These 
values comparable to that of the star-forming region
W49N (Vastel et al.~2001) and are significantly larger than the 
values of G$_0$ in the prototypical and best studied PDRs.

In Fig.~3 we show the CO$^+$ and HOC$^+$ column densities derived in
this paper and in Fuente et al.~(2003), as a function of the incident 
UV field. Error bars represent the source-averaged and beam-averaged 
limits, and hence do not take into account other sources of 
uncertainty. The beam filling factor was computed by taking into account
the angular extension of each PDR. 
The assumed sizes are: 
a filament of 6$''$ of thickness in NGC 7023 (Fuente et al.~1996);
half of the beam in the case of the Orion Bar since the (0$''$, 0$''$) 
position
is at the edge of the molecular bar (Fuente et al.~2003);
13$''$ in NGC 7027 (Cox et al.~1997); and the sizes of the
radio continuum emission at 6cm, 10$''$ and 8$''$ in Mon R2 and \g\
respectively. Mon R2 and \g\ are at the upper end of the range of UV 
fields studied so far. A selection effect may be present in Fig.~3 
because the detection are only possible above a certain column density 
(approximately the upper limit of \hocp\ in NGC 7027). Even so, when 
detected, the column densities remain of the same order in a variety 
of objects, which is graphically shown in the large range of G$_0$ 
in Fig.~3. Although the UV fields are spread over two
orders of magnitude, the CO$^+$ column densities measured towards
Mon R2 and \g\ are closely similar, within a factor of 2, to those observed in
the Orion Bar and NGC 7023. This would imply that the total CO$^+$ column
density in a PDR remain roughly constant for a wide range of global physical
conditions ($n$ from 10$^5$ cm$^{-3}$ to $>$ 10$^6$ cm$^{-3}$)
and UV radiation fields (G$_0$ from 10$^3$ to $> 10^5$). Only the 
C-rich planetary nebula NGC 7027 seems rather different from the 
rest of sources. A similar behavior is found for HOC$^+$.

The chemistry of CO$^+$ is still poorly known.
Several CO$^+$ formation mechanisms, including
C$^+$ + OH $\rightarrow$ CO$^+$ + H, the charge transfer reaction
between CO and CH$^+$, and the direct ionization of CO, have
been proposed to explain the large \cop\ abundances observed. 
The almost constant CO$^+$ abundance argues in favour of the direct
photoionization of CO is not a significant formation mechanism for CO$^+$
(Fuente \& Mart{\i}n-Pintado~1997). Models in which only the
C$^+$ + OH 
and CH$^+$ + CO reactions are included (Sternberg \& Dalgarno 1995;
Hasegawa et al.\ 2000) accounts reasonably well for the CO$^+$ column
densities plotted in Fig. 3. 

On the other hand, HOC$^+$ is efficiently formed in a PDR {\it via} 
the reaction
C$^+$ + H$_2$O $\rightarrow$ HOC$^+$/HCO$^+$ + H, and
CO$^+$ + H$_2$ $\rightarrow$ HOC$^+$/HCO$^+$ + H, but it is
rapidly destroyed by the isomerization reaction with H$_2$ 
(Smith et al.~2002). 
The low values of [HCO$^+$]/[HOC$^+$] observed in PDRs
can only be explained by a very rapid loss of the isomer HCO$^+$ by 
electronic recombination in regions with high electron density. There
is no detection of HOC$^+$ in NGC 7027, which is quite surprising 
regarding the large CO$^+$ column
density measured in this source. One possible explanation is that the
bulk of the HOC$^+$ and HCO$^+$ in PDRs are formed by
the reaction C$^+$ + H$_2$O in a region where the CO$^+$ is
not abundant. This path is reinforced by the low abundance of water 
in NGC 7027 (Liu et al.~1996). In order to disentangle among 
the different processes, detailed chemical models and 
interferometric observations --which will trace the relative positions 
of the CO$^+$ and HOC$^+$ layers-- are required.


%% Included in this acknowledgments section are examples of the
%% AASTeX hypertext markup commands. Use \url without the optional [HREF]
%% argument when you want to print the url directly in the text. Otherwise,
%% use either \url or \anchor, with the HREF as the first argument and the
%% text to be printed in the second.

\acknowledgments
We are grateful to the technical staff of Pico de Veleta for their kindly
support during the observations.
This paper has been partially funded by the Spanish MCyT
under projects DGES/AYA2000-927, AYA 2002-10113-E, ESP2001-4519-PE, 
ESP2002-01693 and ESP2002-01627.

%%%%%%%%%%%%%%%%%%%%%%%%%%%%%%%%%%%%%%%%%%%%%%%%%%%%%%%%%%%%%%%%%%%%%%%%%%%%%%%%%%
%%%%%%%%%%%%%%     Table 1        %%%%%%%%%%%%%%%%%

\clearpage

\begin{table}
%\tablewidth{0pt}
\begin{center}
\caption{Observing frequencies and telescope parameters}
\begin{tabular}{llcc}
\tableline\tableline
\multicolumn{1}{l}{Line} & \multicolumn{1}{l}{Freq (GHz)} &
\multicolumn{1}{c}{beam ($''$)} & \multicolumn{1}{c}{$\eta_{MB}$} \\
\tableline
HOC$^+$ 1 $\rightarrow$ 0         &  89.4874 & 27.5 & 0.77  \\
H$^{13}$CO$^+$ 1 $\rightarrow$ 0  &  86.7543 & 28 & 0.78   \\
HC$^{18}$O$^+$ 1 $\rightarrow$ 0  &  85.1622 & 29 & 0.78   \\
CO$^+$  2,5/2 $\rightarrow$ 1,3/2 & 236.0625 & 10.5 & 0.51  \\
CO$^+$  2,3/2 $\rightarrow$ 1,1/2 & 235.7896 & 10.5 & 0.51  \\
%SO$^+$  5/2 $\rightarrow$ 3/2 (e) & 115.804405 & --- & --- \\
%SO$^+$  5/2 $\rightarrow$ 3/2 (f) & 116.179947 & --- & --- \\
%HCNH$^+$ 2$\rightarrow$1          & 148.2214 & --- & --- \\
%HCNH$^+$ 3$\rightarrow$2          & 222.3294 & --- & --- \\
%C$_3$H$_2$ 3(2,2) $\rightarrow$ 2(1,1)  & 155.5183   & 17$''$  & 0.65  \\
%C$_3$H$_2$ 6(0,6) $\rightarrow$ 5(1,5)  & 217.8220  & 12$''$  & 0.54 \\
%C$_3$H$_2$ 6(1,6) $\rightarrow$ 5(0,5)  & 217.8222  & 12$''$  & 0.54 \\
%HC$_3$N  10 $\rightarrow$ 9             &  90.9789933 & --- & --- \\
%HC$_3$N  10 $\rightarrow$ 9 $\nu_7=1$   &  91.2026 & --- & --- \\
%HCO 1 $\rightarrow$ 0             &  86.6708200 & 29$''$ & 0.80   \\
%HCO 3 $\rightarrow$ 2             & 238.668347  & 10.5$''$ & 0.50  \\
SiO 2 $\rightarrow$ 1             &  86.8468910 & 28 & 0.78   \\
SiO 3 $\rightarrow$ 2             & 130.268702 & 19 & 0.72 \\
\tableline
\end{tabular}
\end{center}
\end{table}

\clearpage
%%%%%%%%%%%%%%%%%%%%%%%%%%%%%%%%%%%%%%%%%%%%%%%%%%%%%%%%%%%%%%%%%%%%%%%%%%%%%%%%%%
%%%%%%%%%%%%%%     Table 2        %%%%%%%%%%%%%%%%%

\begin{table}
%\begin{center}
\begin{scriptsize}
\caption{Observational parameters}
\begin{tabular}{llccc}
\tableline\tableline
\multicolumn{1}{l}{Position} &
\multicolumn{1}{l}{Line} &
%\multicolumn{1}{c}{T$_{\rm MB}$} &
\multicolumn{1}{c}{V$_{\rm LSR}$} &
\multicolumn{1}{c}{$\Delta$V} &
\multicolumn{1}{c}{Area} \\
\multicolumn{1}{l}{($''$,$''$)} &
\multicolumn{1}{c}{} &
%\multicolumn{1}{c}{(K)} &
\multicolumn{2}{c}{(km s$^{-1}$)} &
\multicolumn{1}{c}{(K km s$^{-1}$)} 
\\
\tableline
Mon R2 & & & &\\
(0,0)     & HC$^{18}$O$^+$ 1$\rightarrow$0 &  9.6 (4) & 1.3 (8) & 0.07 (4) \\
          & H$^{13}$CO$^+$ 1$\rightarrow$0 & 10.5 (5) & 5.4 (8) & 0.57 (9) \\
          & HOC$^+$ 1$\rightarrow$0        &  9.9 (8) & 2.2 (9) & 0.06 (4) \\
          & CO$^+$ 2,5/2$\rightarrow$1,3/2 &  9.7 (7) & 4.3 (9) & 0.46 (9) \\
          & CO$^+$ 2,3/2$\rightarrow$1,1/2 & 10.4 (5) & 2.8 (9) & 0.27 (9) \\
          & SiO 2$\rightarrow$1            &  9.6 (3) & 1.1 (9) & 0.05 (2) \\
          & SiO 3$\rightarrow$2            &  9.7 (7) & 3.0 (9) & 0.26 (9) \\
\\
(+10,-10) & HC$^{18}$O$^+$ 1$\rightarrow$0 & 10.7 (5) & 1.7 (9) & 0.11 (6) \\
          & H$^{13}$CO$^+$ 1$\rightarrow$0 & 11.2 (9) & 3.5 (9) & 1.4  (4) \\
          & HOC$^+$ 1$\rightarrow$0        & --- & --- & $< 0.03$ \\
          & CO$^+$ 2,5/2$\rightarrow$1,3/2 & --- & --- & $< 0.08$ \\
          & CO$^+$ 2,3/2$\rightarrow$1,1/2 & --- & --- & $< 0.08$ \\
\\
\multicolumn{3}{l}{G29.96-0.02}\\
(0,0)     & HC$^{18}$O$^+$ 1$\rightarrow$0 &  97.3 (1) &  2.9 (3) & 0.95 (6) \\
          & H$^{13}$CO$^+$ 1$\rightarrow$0\tablenotemark{(W)} &  96.2 (9) & 16.1 (9) & 1.0  (4) \\
          & H$^{13}$CO$^+$ 1$\rightarrow$0\tablenotemark{(N)} &  97.7 (1) &  4.4 (2) & 6.7  (3) \\
          & HOC$^+$ 1$\rightarrow$0        &  96.1 (9) &  3.0 (9) & 0.08 (5) \\
          & CO$^+$ 2,5/2$\rightarrow$1,3/2\tablenotemark{(B)} &  98.1 (9) &  7.1 (9) & 3.79 (9) \\
          & CO$^+$ 2,3/2$\rightarrow$1,1/2\tablenotemark{(M)} &  98.0 (4) &  1.3 (9) & 0.14 (8) \\
          & CO$^+$ 2,3/2$\rightarrow$1,1/2\tablenotemark{(R)} & 101.3 (3) &  1.0 (8) & 0.10 (7) \\
          & SiO 2$\rightarrow$1\tablenotemark{(W)}            &  97.3 (1) & 18.9 (9) & 3.6  (1) \\
          & SiO 2$\rightarrow$1\tablenotemark{(N)}            &  97.5 (1) &  4.6 (2) & 2.3  (1) \\
          & SiO 3$\rightarrow$2\tablenotemark{(W)}            &  97.7 (1) &  5.0 (3) & 2.8  (3) \\
          & SiO 3$\rightarrow$2\tablenotemark{(N)}            &  97.9 (3) & 18.6 (9) & 5.2  (3) \\
\tableline
\end{tabular}
\tablenotetext{}{Numbers in parenthesis are 3$\sigma$ uncertainties in 
the last digit.}
\tablenotetext{W}{Wide component}
\tablenotetext{N}{Narrow component}
\tablenotetext{B}{Blended with $^{13}$CH$_3$OH}
\tablenotetext{M}{Main component}
\tablenotetext{R}{Redshifted component}

\end{scriptsize}
%\end{center}
\end{table}

%%%%%%%%%%%%%%     Table 3        %%%%%%%%%%%%%%%%%

\clearpage

\begin{table}
%\begin{center}
\caption{Beam-averaged column densities (in cm$^{-2}$)}
\begin{tabular}{lccccc}
\tableline\tableline
\multicolumn{1}{l}{Position} &
\multicolumn{1}{c}{SiO} &
\multicolumn{1}{c}{HC$^{18}$O$^+$} &
\multicolumn{1}{c}{H$^{13}$CO$^+$} &
\multicolumn{1}{c}{HOC$^+$} &
\multicolumn{1}{c}{CO$^+$} \\

%\multicolumn{1}{l}{} &
%\multicolumn{1}{c}{cm$^{-2}$} &
%\multicolumn{1}{c}{cm$^{-2}$} &
%\multicolumn{1}{c}{cm$^{-2}$} &
%\multicolumn{1}{c}{cm$^{-2}$} &
%\multicolumn{1}{c}{cm$^{-2}$} \\
\tableline

Mon R2 & & & & & \\
(0,0)     & 1.0 10$^{11}$ & 1.2 10$^{11}$ & 1.1 10$^{12}$ & 1.7 10$^{11}$ & 5.3 10$^{11}$ \\
(+10,-10) & ---     & 1.8 10$^{11}$ & 2.6 10$^{12}$ & $<$0.8 10$^{11}$ & $<$1.6 10$^{11}$ \\
\\
\multicolumn{3}{l}{G29.96-0.02}  & &\\
(0,0)wide & 1.1 10$^{13}$ & ---     & 1.5 10$^{12}$ & ---        & ---    \\
(0,0)narr.& 6.9 10$^{12}$ & 1.5 10$^{12}$ & 9.0 10$^{12}$ & 2.3 10$^{11}$    & 4.7 10$^{11}$ \\

\tableline
\end{tabular}

%\end{center}
\end{table}

%%%%%%%%%%%%%%%%%%%%%%%%%%%%%%%%%%%%%%%%%%%%%%%%%%%%%%%%%%%%%%%%%%%%%%%%%%%%

\clearpage


\begin{figure*}
%\epsscale{0.7}
\plotone{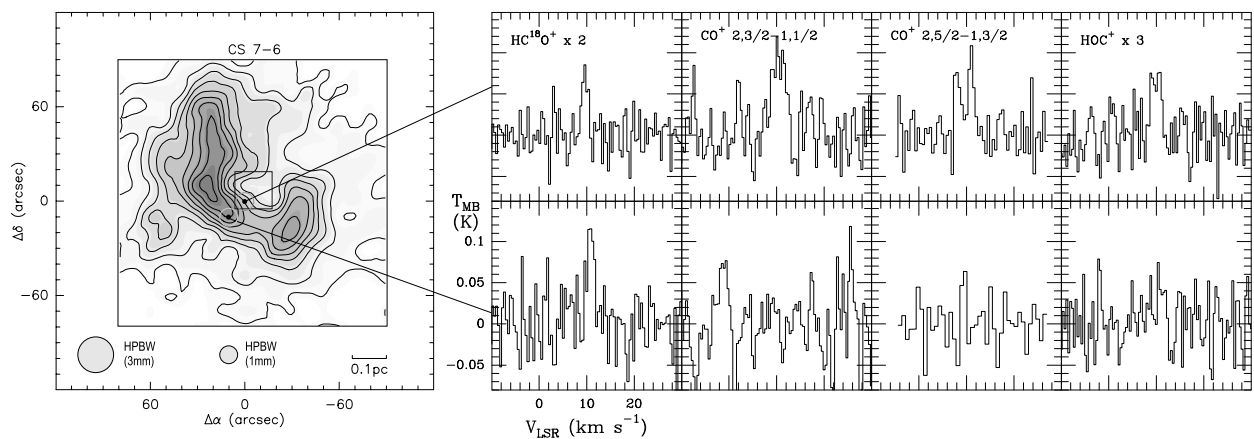}
\caption{{\bf Left panel.} CS 7-6 emission map of Mon R2, taken from
Choi et al.~(2000). Offsets are referred to
$(\alpha, \delta)_{2000} = (6^{\rm h}07^{\rm m}46\fs 2, -6\degr
23\arcmin 08\farcs 3)$. The \uch\ is indicated by the central square. 
The two positions observed in this work are marked by dots.
{\bf Right panel.} Molecular ions observed at the \HII\ region and
its molecular envelope in Mon R2. It is remarkable the detection of 
\cop\ and \hocp\ toward the \uch\ and the non-detection of
the same lines at the molecular cloud.
\label{fig1}}
\end{figure*}

\clearpage

\begin{figure}
\epsscale{0.6}
\plotone{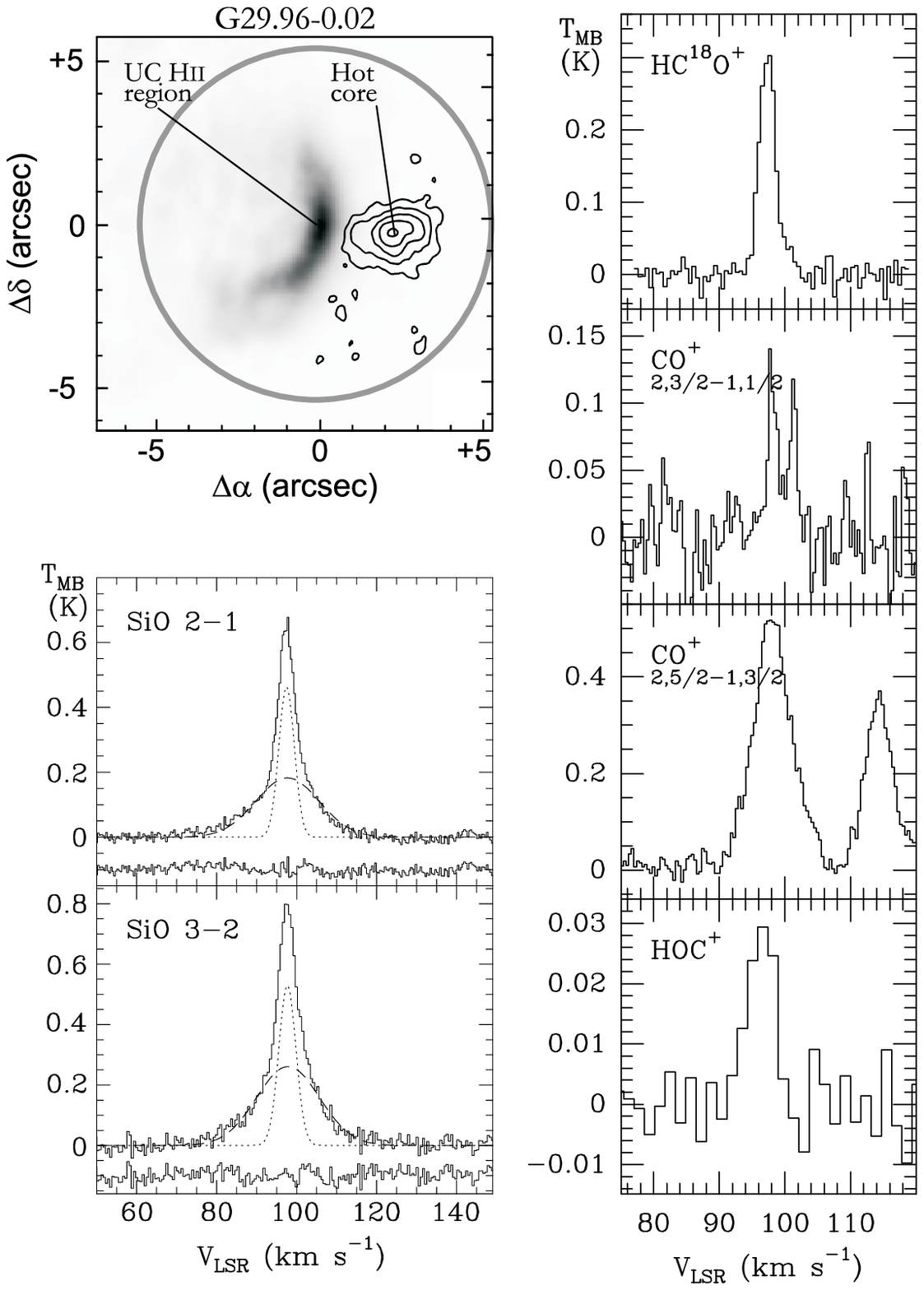}
\caption{{\bf Upper-left panel.} Relative positions of the \uch\ \g\
and its neighbour hot core. The 1.3 cm-continuum emission is shown in
grey, and the NH$_3$ (4,4) emission is marked by contours (Cesaroni
et al.~1998). Offsets are referred to
$(\alpha, \delta)_{2000} = (18^{\rm h}46^{\rm m}03\fs 9,
-2\degr 39\arcmin 21\farcm 9)$.
The large circle enlarges the 1mm beam of our observations.
{\bf Lower-left panel.} SiO 2--1 and 3--2 spectra and gaussian
fits, showing the
contribution of the hot core (wide component) and the molecular 
envelope (narrow component). The residual spectra are 
also plotted.
{\bf Right panel.} Molecular ions observed at \g. It is noticeable the
asymmetry in the HC$^{18}$O$^+$, the double-peaked structure of the
2,3/2 $\rightarrow$ 1,1/2 line of \cop\ and the blending of the 
2,5/2 $\rightarrow$ 1,3/2 line of \cop\ with $^{13}$CH$_3$OH.
\label{fig2}}
\end{figure}

\clearpage

\begin{figure}
\epsscale{0.7}
\plotone{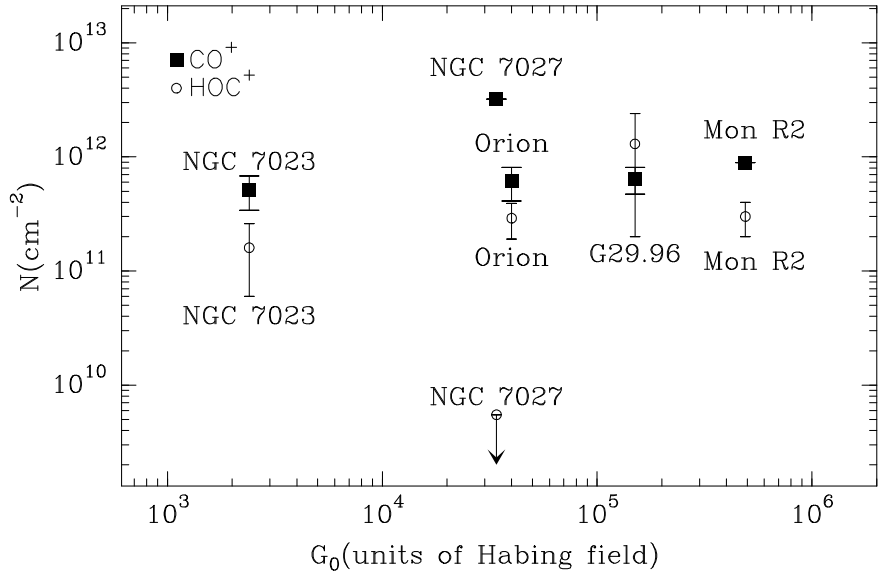}
\caption{
Plot of the CO$^+$ (black squares) and HOC$^+$ (empty circles) column densities
as a function of the incident UV field, for the PDRs reported by Fuente et
al.~(2003) and this paper. Error bars in column densities
represent the lower and upper limits, determined by beam-averaged and 
source-averaged estimates.}
\label{fig3}
\end{figure}

%% If you are not including electonic art with your submission, you may
%% mark up your captions using the \figcaption command. See the
%% No more than seven \figcaption commands are allowed per page,
%% so if you have more than seven captions, insert a \clearpage

\end{document}